\documentclass[twoside,twocolumn,9pt]{article}
\usepackage{extsizes}
\usepackage[super,sort&compress,comma]{natbib} 
\usepackage[version=3]{mhchem}
\usepackage[left=1.5cm, right=1.5cm, top=1.785cm, bottom=2.0cm]{geometry}
\usepackage{balance}
\usepackage{mathptmx}
\usepackage{sectsty}
\usepackage{graphicx} 
\usepackage{lastpage}
\usepackage[format=plain,justification=justified,singlelinecheck=false,font={stretch=1.125,small,sf},labelfont=bf,labelsep=space]{caption}
\usepackage{float}
\usepackage{fancyhdr}
\usepackage{fnpos}
\usepackage[english]{babel}
\addto{\captionsenglish}{%
  
}
\usepackage{array}
\usepackage{droidsans}
\usepackage{charter}
\usepackage[T1]{fontenc}
\usepackage[usenames,dvipsnames]{xcolor}
\usepackage{setspace}
\usepackage[compact]{titlesec}
\usepackage{hyperref}
\usepackage{epstopdf}%This line makes .eps figures into .pdf - please comment out if not required.

\definecolor{cream}{RGB}{222,217,201}

\begin{document}

\pagestyle{fancy}
\thispagestyle{plain}
\fancypagestyle{plain}{
%%%HEADER%%%
\renewcommand{\headrulewidth}{0pt}
}
%%%END OF HEADER%%%

%%%PAGE SETUP - Please do not change any commands within this section%%%
\makeFNbottom
\makeatletter
\renewcommand\LARGE{\@setfontsize\LARGE{15pt}{17}}
\renewcommand\Large{\@setfontsize\Large{12pt}{14}}
\renewcommand\large{\@setfontsize\large{10pt}{12}}
\renewcommand\footnotesize{\@setfontsize\footnotesize{7pt}{10}}
\makeatother

\renewcommand{\thefootnote}{\fnsymbol{footnote}}
\renewcommand\footnoterule{\vspace*{1pt}% 
\color{cream}\hrule width 3.5in height 0.4pt \color{black}\vspace*{5pt}} 
\setcounter{secnumdepth}{5}

\makeatletter 
\renewcommand\@biblabel[1]{#1}            
\renewcommand\@makefntext[1]% 
{\noindent\makebox[0pt][r]{\@thefnmark\,}#1}
\makeatother 
\renewcommand{\figurename}{\small{Fig.}~}
\sectionfont{\sffamily\Large}
\subsectionfont{\normalsize}
\subsubsectionfont{\bf}
\setstretch{1.125} %In particular, please do not alter this line.
\setlength{\skip\footins}{0.8cm}
\setlength{\footnotesep}{0.25cm}
\setlength{\jot}{10pt}
\titlespacing*{\section}{0pt}{4pt}{4pt}
\titlespacing*{\subsection}{0pt}{15pt}{1pt}
%%%END OF PAGE SETUP%%%

%%%FOOTER%%%
\fancyfoot{}
\fancyfoot[LO,RE]{\vspace{-7.1pt}\includegraphics[height=9pt]{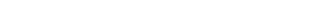}}
\fancyfoot[CO]{\vspace{-7.1pt}\hspace{13.2cm}\includegraphics{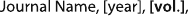}}
\fancyfoot[CE]{\vspace{-7.2pt}\hspace{-14.2cm}\includegraphics{head_foot/RF}}
\fancyfoot[RO]{\footnotesize{\sffamily{1--\pageref{LastPage} ~\textbar  \hspace{2pt}\thepage}}}
\fancyfoot[LE]{\footnotesize{\sffamily{\thepage~\textbar\hspace{3.45cm} 1--\pageref{LastPage}}}}
\fancyhead{}
\renewcommand{\headrulewidth}{0pt} 
\renewcommand{\footrulewidth}{0pt}
\setlength{\arrayrulewidth}{1pt}
\setlength{\columnsep}{6.5mm}
\setlength\bibsep{1pt}
%%%END OF FOOTER%%%

%%%FIGURE SETUP - please do not change any commands within this section%%%
\makeatletter 
\newlength{\figrulesep} 
\setlength{\figrulesep}{0.5\textfloatsep} 

\newcommand{\topfigrule}{\vspace*{-1pt}% 
\noindent{\color{cream}\rule[-\figrulesep]{\columnwidth}{1.5pt}} }

\newcommand{\botfigrule}{\vspace*{-2pt}% 
\noindent{\color{cream}\rule[\figrulesep]{\columnwidth}{1.5pt}} }

\newcommand{\dblfigrule}{\vspace*{-1pt}% 
\noindent{\color{cream}\rule[-\figrulesep]{\textwidth}{1.5pt}} }

\makeatother
%%%END OF FIGURE SETUP%%%

%%%TITLE, AUTHORS AND ABSTRACT%%%
\twocolumn[
  \begin{@twocolumnfalse}
{\includegraphics[height=30pt]{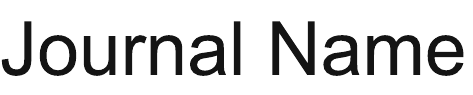}\hfill\raisebox{0pt}[0pt][0pt]{\includegraphics[height=55pt]{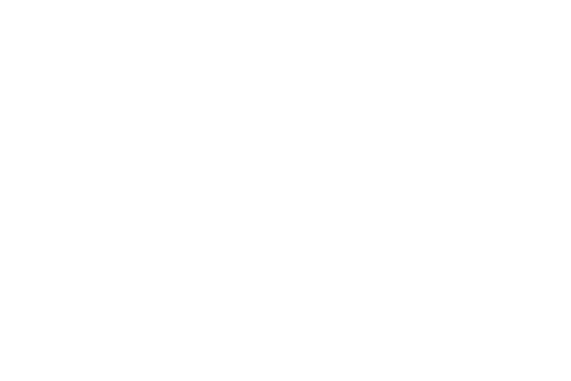}}\\[1ex]
\includegraphics[width=18.5cm]{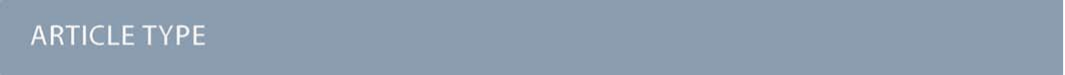}}\par
\vspace{1em}
\sffamily
\begin{tabular}{m{4.5cm} p{13.5cm} }

\includegraphics{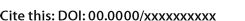} & \noindent\LARGE{\textbf{Metastable \ce{Cu_{1--$x$}CrTe2} -- Completing the copper chromium delafossite series through soft chemistry}} \\%Article title goes here instead of the text "This is the title"
\vspace{0.3cm} & \vspace{0.3cm} \\

 & \noindent\large{Kai D. Röseler,\textit{$^{a}$} Geo Sciarini,\textit{$^{a}$} Felix Eder,\textit{$^{a}$} Samuel Moody,\textit{$^{b}$} Vladimir Pomjakushin,\textit{$^{b}$} and Fabian O. von Rohr$^{\ast}$\textit{$^{a}$}} \\%Author names go here instead of "Full name", etc.

\includegraphics{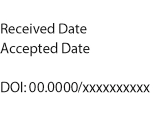} & \noindent\normalsize{%Accessing metastable Cu(I)-containing compounds remains a central challenge in solid-state synthesis because high-temperature routes often favor decomposition into thermodynamically preferred phases. 
In the layered copper chromium dichalcogenide series \ce{CuCr$X$2}, the oxide, the sulfide, and the selenide analogues have been reported, but the telluride \ce{CuCrTe2} has remained unsynthesized so far. Here, we report the synthesis of \ce{Cu_{1--$x$}CrTe2} ($x \approx 0.3$), which forms only within a narrow temperature window near 90\,°C by solvothermal cation exchange of \ce{K_{1--$x$}CrTe2} ($x \approx 0.3$) and CuBr. \ce{Cu_{1--$x$}CrTe2} undergoes a magnetostructural transition to an antiferromagnetic state at $T_\mathrm{N}=239$ K, a Néel temperature that is high relative to other \ce{$A$CrTe2} phases and comparable to that of ferromagnetic, fully-deintercalated \ce{CrTe2}. \ce{Cu_{1--$x$}CrTe2} is metastable and decomposes at temperatures as low as 250\, °C to form spinel \ce{CuCr2Te4}. These results highlight the importance of low-temperature topochemical routes for accessing metastable Cu(I)-containing tellurides that are inaccessible by conventional solid-state synthesis.

}\\

\end{tabular}

 \end{@twocolumnfalse} \vspace{0.6cm}

  ]
%%%END OF TITLE, AUTHORS AND ABSTRACT%%%

%%%FONT SETUP - please do not change any commands within this section
\renewcommand*\rmdefault{bch}\normalfont\upshape
\rmfamily
\section*{}
\vspace{-1cm}

%%%FOOTNOTES%%%

\footnotetext{\textit{$^{a}$~Department of Quantum Matter Physics, University of Geneva, CH-1211 Geneva, Switzerland.}}

\footnotetext{\textit{$^{b}$~Laboratory for Neutron Scattering and Imaging, Paul Scherrer Institute, CH-5232 Villigen PSI, Switzerland}}

%Please use \dag to cite the SI in the main text of the article.
%If you article does not have SI please remove the the \dag symbol from the title and the footnotetext below.
%\footnotetext{\dag~Supplementary Information available: [details of any supplementary information available should be included here]. See DOI: 00.0000/00000000.}
%additional addresses can be cited as above using the lower-case letters, c, d, e... If all authors are from the same address, no letter is required

%\footnotetext{\dag~Additional footnotes to the title and authors can be included \textit{e.g.}\ `Present address:' or `These authors contributed equally to this work' as above using the symbols: \ddag, \textsection, and \P. Please place the appropriate symbol next to the author's name and include a \texttt{\textbackslash footnotetext} entry in the the correct place in the list.}

%%%END OF FOOTNOTES%%%

%%%MAIN TEXT%%%%
\section*{Introduction}

The discovery of new inorganic solids is generally limited because conventional high-temperature synthesis bypasses metastable phases in favor of thermodynamically preferred products.\cite{jansen2002concept,Cordova2020_metastable} Topochemical reactions, which preserve structural motifs, offer a way around this limitation by operating at low temperatures. They are especially powerful for layered compounds where guest species in the interlayer space can be intercalated,\cite{Meyer_2010_OrgInt,Kelly_2025_LiInt,Hatakeda_2013_Li_en_Int,Hyde_2024_LiKInt,Hyde_2023_LiInt,witte2024tuning,tezze2026galvanic} removed,\cite{Freitas2015,Roeseler2025_CrTe2,Zhou_2016_CoSe,mantravadi2024van} or exchanged.\cite{Schoop_2025_CuEx,Sethi_Excahnge,song2025transitory} A striking example is 1T-\ce{CrTe2}, a van der Waals ferromagnet with a Curie temperature above room temperature, which is accessible only through topochemical deintercalation of \ce{$A$CrTe2} ($A$ = Li, K).\cite{Freitas2015,Roeseler2025_CrTe2}

\ce{Cr$X$2} ($X$ = chalcogen) compounds consist of triangular lattices of Cr atoms arranged in layered chalcogenide slabs separated by van der Waals gaps.\cite{Freitas2015,song2019soft,van1980crse2} These gaps can accommodate interlayer cations, providing access to the broader \ce{$A$Cr$X$2} family.\cite{engelsman1973crystal,roseler2026ladder,song2021kinetics,eder2025structural,eder2025stoichiometry} In these phases, both the interlayer ion and the chalcogen strongly influence magnetic exchange, transport, and structural distortions, making \ce{$A$Cr$X$2} compounds a platform for exploring various electronic and magnetic ground states.\cite{kobayashi2016competition,nocerino2023competition,nocerino2022nuclear} Beyond the intercalated $A$ cation, the chalcogen itself strongly influences the physical properties, as exemplified in the \ce{CuCr$X$2} family. Within this family, \ce{CuCrO2} is a transparent p-type semiconductor,\cite{CuCrO2_ref} \ce{CuCrS2} exhibits a remarkably large Seebeck coefficient\cite{Tewari_2010_CuCrS,romanenko2022review}, and single-layer \ce{CuCrSe2} is a promising multiferroic candidate.\cite{Sun_2024_CuCrSe2_Multiferro,Rahman_2025_OrderDisorder} Yet the telluride analogue \ce{CuCrTe2} has not been reported so far. While it has been predicted theoretically to exist as a metallic antiferromagnet.\cite{srivastava2013first} 

Cr-based tellurides pose a particular synthetic challenge because high-temperature reactions favor competing phases, including NiAs-type \ce{Cr_{\textit{x}}Te_{\textit{y}}} compounds such as \ce{Cr2Te3}.\cite{CrxTey_PhaseDiagram} In the Cu--Cr--Te system, an additional thermodynamic sink is provided by the ferromagnetic spinel \ce{CuCr2Te4} ($T_\mathrm{C}=326$\,K),\cite{Suzuyama_2006,koneshova2014ternary} as well as Cu-rich variants \ce{Cu_{1+\textit{y}}Cr2Te4} ($0<y\le1$) with lower Curie temperatures.\cite{Lotgering_1971_excess} Already for the lighter chalcogenides, synthesis of the layered phase requires careful temperature control to avoid spinel formation: \ce{CuCrS2} was synthesized at $650$\,°C to suppress \ce{CuCr2S4} impurities,\cite{Tewari_2010_CuCrS} and \ce{CuCrSe2} is reported to be in equilibrium with \ce{CuCr2Se4} at temperatures as low as $300$\,°C.\cite{CuCrSe2_eq_ref}

In this work, we present the synthesis of \ce{Cu_{1--$x$}CrTe2} ($x \approx 0.3$) using a solvent-mediated cation exchange reaction from \ce{K_{1--$x$}CrTe2} ($x \approx 0.3$) in comparison to conventional solid-state syntheses and self-flux syntheses, which yielded \ce{CuCr2Te4} or metal-excessive \ce{Cu_{1+\textit{y}}Cr2Te4} spinels. We found that \ce{Cu_{1--$x$}CrTe2} decomposes as low as $250$\,°C into the structurally related \ce{CuCr2Te4} spinel. Using a combined analysis of single-crystal X-ray diffraction, magnetization, and neutron powder diffraction experiments, we identified a magneto-structural transition towards an antiferromagnetic state with $T_{\mathrm{N}} = 239$\,K.

\newpage

\section*{Experimental}

\subsection*{Conventional solid-state syntheses towards \ce{CuCrTe2}}
Te (pieces, Alfa Aesar, 99.999\%) was ground in an agate mortar and mixed with Cu powder (Merck, 99.5\%) and Cr powder (Alfa Aesar, 99.95\%) in a molar ratio of 2:1:1. The mixture was then pressed into 500\,mg pellets with a diameter of 8\,mm, placed into quartz tubes (inner diameter 10\,mm, thickness 1\,mm, length $\approx8.5$\,cm), and flame sealed under a 300\,mbar Ar atmosphere. The sealed silica tubes were placed into box furnaces, which were heated at a rate of 50\,K$\cdot$h$^{-1}$ to $T_{x}$, which was held for 120\,h. The maximum temperatures $T$ investigated included 150\,°C, 350\,°C, and 600\,°C.

\subsection*{Self-flux syntheses towards \ce{CuCrTe2}}
Cu (Merck, 99.5\%), Cr (powder, Alfa Aesar, 99.95\%) and Te (powder, ground from pieces, Alfa Aesar, 99.999\%) in molar ratios of $Z$:1:8 ($Z=4,5,6$) were thoroughly mixed and placed in a Canfield alumina crucible set comprising a bottom and top crucible and a frit separating disc in between \cite{Canfield2016}. The set was then sealed in a quartz ampule (inner diameter 12\,mm, thickness 1.5\,mm, length $\approx11.5$\,cm) under an Ar atmosphere of 300\,mbar. Inside a muffle furnace, the ampule was heated at 30\,°C$\cdot$h$^{-1}$ to 1000\,°C, and was subsequently slow-cooled to 750\,°C within 96\,h. At 750\,°C, the ampule was removed from the oven, and the excess flux was separated by immediate high-temperature centrifugation.

\subsection*{Solvothermal cation exchange reactions towards \ce{CuCrTe2}}
Solvothermal cation exchange reactions were performed using CuBr as a Cu\textsuperscript{+} source and three different $A$\ce{CrTe2} phases: (1) Solid-state synthesized \ce{K_{1--\textit{x}}CrTe2} ($x \approx 0.3$), (2) flux-grown \ce{K_{1--\textit{x}}CrTe2} ($x \approx 0.3$), and (3) flux-grown \ce{LiCrTe2}. Inside an Ar-filled glovebox, 50\,mg of the respective $A$\ce{CrTe2} phases and CuBr (molar ratio 1:10) were placed in Teflon containers with an inner volume of about 16\,ml. After the addition of 7\,ml dry acetonitrile, the container was closed with a Teflon lid and mounted into a steel autoclave. The autoclave was then heated to 90\,°C or 200\,°C and it was kept for 7 days under autogenous pressure. Afterward, the oven was turned off, and the autoclave was cooled to room temperature. The autoclave was opened under an Ar atmosphere, and the crystals were transferred into Schlenk tubes. Using standard Schlenk techniques, excess CuBr and KBr were removed by washing the crystals three times with 7\,ml of dry acetonitrile, and the crystals were dried under reduced pressure. When increasing the amount of $A$\ce{CrTe2} to 100\,mg, Teflon containers with an inner volume of 44\,ml and 14\,ml of dry acetonitrile were used instead.

\subsection*{Precursor syntheses}
Self-flux syntheses of \ce{LiCrTe2} and \ce{K_{1--$x$}CrTe2} ($x \approx 0.3$) were conducted as previously reported.\cite{witteveen2023synthesis,Witteveen_2025} For the solid-state synthesis of \ce{K_{1--$x$}CrTe2} ($x \approx 0.3$), Cr (powder, Alfa Aesar, 99.95\%), Te (pieces, Alfa Aesar, 99.999\%), and K (block, CALLERY, min 99\,\%) in a molar ratio of 1:1:2 were placed in an alumina crucible. The crucible and an additional crucible on top were flame-sealed inside a quartz tube (inner diameter 14\,mm, thickness 2\,mm, length 8\,cm). The quartz tube was then placed inside a box furnace, which was heated at a rate of 30\,Kh$^{-1}$ to 900\,°C and held at this temperature for 8 days. The furnace was then cooled at 6\,Kh$^{-1}$ to 450\,°C and subsequently at 60\,Kh$^{-1}$ to room temperature.

\subsection*{Powder X-ray diffraction (PXRD)}
PXRD data were collected with a Rigaku SmartLabXE diffractometer with Cu-K\textsubscript{$\alpha$} radiation ($\lambda$\,=\,1.54187\,\AA) on a D/teX Ultra 250 detector in capillary mode in the 2$\theta$ range of 5--80°. Crystals were ground into fine powders, mixed with starch to minimized effects of preferred orientation, and filled into borosilicate capillaries with an outer diameter of 0.8\,mm. Rietveld refinements of the intensity data was performed in the TOPAS-Academic software.\cite{Topas}

\subsection*{Single X-ray diffraction (SXRD)} 
Single X-ray diffraction (SXRD) experiments were performed under \ce{N2} cooling at 100\,K, 180\,K, and 293\,K on an Oxford Diffraction Supernova diffractometer using Mo K\textsubscript{$\alpha$} radiation ($\lambda$\,=\,0.71073\,\AA). Pre-experiment screenings, data collection, data reduction, and absorption correction were performed using the program suite CrysAlisPro.\cite{Rigaku2015} The crystal structure was solved with the dual space method in SHELXT.\cite{Sheldrick2015XT} Least squares refinements of $F^2$ were performed using SHELXL.\cite{Sheldrick2015XL}

\subsection*{Scanning electron microscopy (SEM) and Energy-dispersive X-ray spectroscopy (EDS)} 
Electron images were obtained from a JEOL JSM-IT800 Scanning electron microscope. Energy dispersive X-ray spectroscopy (EDS) data was collected with an acceleration voltage of 20\,kV with an X-Max\textsuperscript{N} 80 detector from Oxford Instruments. Stoichiometry calculations of \ce{Cu_{1--$x$}CrTe2} ($x \approx 0.3$) from flux-grown \ce{K_{1--$x$}CrTe2} ($x \approx 0.3$) crystals are based on 80 point scans on two sites of four crystals from two batches. The same number of measurements was repeated after exfoliation with Scotch tape. Stoichiometry calculations of solid-state synthesized \ce{K_{1--$x$}CrTe2} ($x \approx 0.3$) and \ce{Cu_{1--$x$}CrTe2} ($x \approx 0.3$) obtained from it are based on 80 point scans. Before the EDS measurements, solid-state synthesized \ce{K_{1--$x$}CrTe2} ($x \approx 0.3$) crystals were washed with dry DMF to remove possible potassium-polytelluride side-products on the surface.

\subsection*{Magnetization experiments} 
Temperature- and field-dependent magnetization measurements were carried out in a Physical Property Measurement System in a cryogen-free system (PPMS DynaCool) from Quantum Design equipped with the vibrating sample magnetometer (VSM) option. \textit{m}(\textit{T}) data were recorded in the 1.8--300\,K or 1.8--360\,K temperature range with a sweep rate of 2\,Km$\cdot$in$^{-1}$, for \textit{m}(H), an interval of $-$9 to 9\,T was scanned at 50\,Oe$\cdot$s$^{-1}$.

\subsection*{Neutron powder diffraction}
Neutron powder diffraction experiments were performed on the DMC Cold Neutron Diffractometer at the Swiss Spallation Neutron Source from the Paul Scherrer Institute in Villigen, Switzerland. Inside a helium-filled glovebox, finely ground powder of \ce{Cu_{1--$x$}CrTe2} ($x \approx 0.3$) obtained from flux-grown \ce{K_{1--$x$}CrTe2} ($x \approx 0.3$) was sealed in an aluminum sample container with an inner diameter of 7.5\,mm using indium wire. Diffraction data were collected at $T=2$\,K, $T=180$\,K, and $T=240$\,K with a neutron wavelength of 2.449\,\AA. Rietveld refinements were performed using the FullProf Suite package, and the magnetic symmetry was analyzed using ISODISTORT in the ISOTROPY software.\cite{Isodistort,Isodisplace}

\subsection*{Differential scanning calorimetry (DSC)}
Differential scanning calorimetry (DSC) measurement was performed using a Mettler Toledo DSC1 STARe System. Ground \ce{Cu_{1--$x$}CrTe2} obtained from solid-state synthesized \ce{K_{1--$x$}CrTe2} ($x \approx 0.3$) was placed in an aluminum container, closed with a lid with a small hole for pressure adjustment and heated at a rate of 5\,Kmin$^{-1}$ under \ce{N2} atmosphere. The temperature program consisted of two cycles: (1) heating and cooling between RT and 150\,°C, and (2) heating and cooling between RT and 300\,°C.

%\subsection*{Tempering of \ce{Cu_{1--$x$}CrTe2} ($x \approx 0.3$)}
%\ce{Cu_{1--$x$}CrTe2} ($x \approx 0.3$)crystals obtained from solvothermal cation exchange reactions from \ce{K_{0.72}CrTe2} were ground to a powder and filled into an alumina crucible. The crucible, together with an empty crucible on top, was sealed in a quartz ampule under 300\,mbar Ar atmosphere. The ampule was then heated at 50\,Kh$^{-1}$ to 250°C, which was held for 48\,h followed by cooling at 50\,Kh$^{-1}$ to RT.

\section*{Results and Discussion}

\subsection*{Synthesis}

\begin{figure*}
 \centering
 \includegraphics[width=0.98\textwidth]{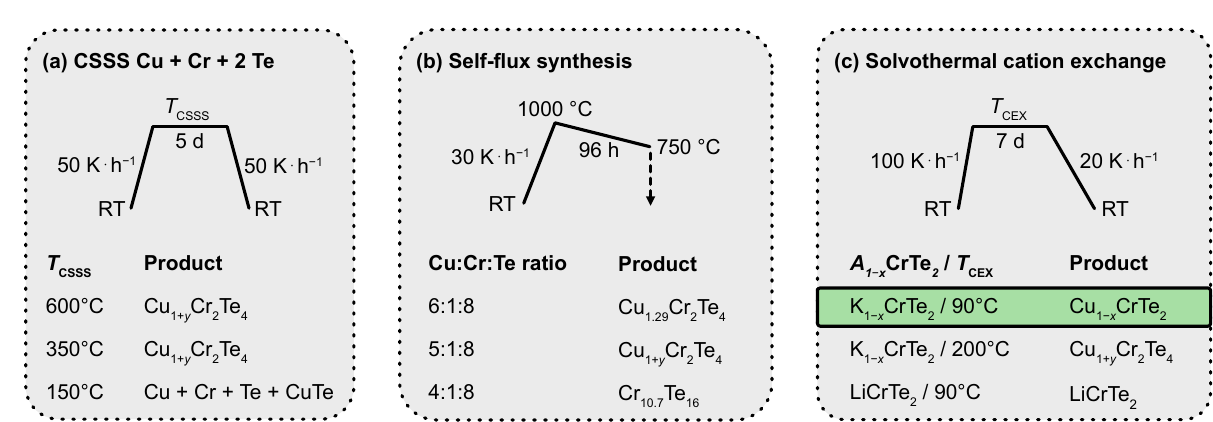}
 \caption{Overview of attempted syntheses towards \ce{CuCrTe2} and main products obtained. (a) Conventional solid-state synthesis (CSSS), (b) self-flux synthesis, and (c) solvothermal cation exchange reactions. Pictograms depict the temperature program used. In (b), a dashed line represents the centrifugation process. In (c), green color highlights the only successful synthesis of \ce{Cu_{1--$x$}CrTe2} ($x \approx 0.3$).}
 \label{fgr:Synth_Overview}
\end{figure*}

We explored three synthetic routes towards \ce{CuCrTe2}: conventional solid-state synthesis (CSSS) from the elements, Cu/Te metal-flux reactions, and solvothermal cation exchange. For the latter, three different precursors were used: solid-state synthesized \ce{K_{1--$x$}CrTe2} ($x \approx 0.3$), flux-grown \ce{K_{1--$x$}CrTe2} ($x \approx 0.3$) crystals, and flux-grown \ce{LiCrTe2} crystals. The reaction conditions tested and products obtained are summarized in Figure \ref{fgr:Synth_Overview}. Solvothermal cation exchange, of solid-state synthesized and flux-grown \ce{K_{1--$x$}CrTe2} ($x \approx 0.3$), at 90\,°C yielded the target phase; all other routes produced \ce{CuCr2Te4} or Cu-rich spinel variants. 

\textbf{Conventional solid-state synthesis (CSSS):} 
Exemplary conventional solid-state syntheses towards \ce{CuCrTe2} were attempted from the elements at 150\,°C, 350\,°C, and 600\,°\,C and the products in the form of black powders were characterized using PXRD and temperature-dependent measurements of the magnetic moment, the results of which are depicted in the SI. After the reaction at 150\,°C, the PXRD of the product shows the presence of the unreacted elements and CuTe. At 350\,°C, we observed a \ce{Cu_{1+$y$}Cr2Te4} spinel as well as four reflections, which we could not attribute to any known phase, but did not correspond to the presence of delafossite-type \ce{CuCrTe2}. The PXRD pattern of the sample synthesized at 600\,°C indicates only the presence of the \ce{Cu_{1+$y$}Cr2Te4} spinel. Measurements of the magnetic moment versus temperature show no magnetic transition between 1.8\,K and 360\,K for the sample obtained at 150\,°C. The sample synthesized at 350\,°C exhibits a ferromagnetic transition at $T_{\mathrm{C}} = 273$\,K and the sample synthesized at 600\,°C a ferromagnetic transition at $T_{\mathrm{C}} = 163$\,K. Both transition temperatures are well below the reported ferromagnetism below $T_{\mathrm{C}} = 326$ \,K \cite{Suzuyama_2006} for \ce{CuCr2Te4}, but within the range of transition temperatures observed for metal-excessive \ce{Cu_{1+\textit{y}}Cr2Te4} ($0<y\le1$) spinels.\cite{Lotgering_1971_excess} Furthermore, our results suggest an increase in excess Cu with elevating tempering temperature, which is also in line with an elongation of the $a$ lattice parameter from 11.2175(4)\,\AA\,($T$\textsubscript{synth}=350\,°C) to 11.3203(4)\,\AA\,($T$\textsubscript{synth}=600\,°C). No single crystals suitable for SXRD experiments were obtained from these syntheses, but Rietveld refinements using a structural model of a copper-rich \ce{Cu_{1+\textit{y}}Cr2Te4} could be performed based on the results from self-flux syntheses discussed in the following.

\textbf{Self-flux synthesis attempts towards \ce{CuCrTe2}:}
Cu/Te self-flux reactions of molar Cu:Cr:Te ratios of 4:1:8, 5:1:8, and 6:1:8 yielded metallic-black crystals. SXRD on the latter identified the product as a copper-rich spinel with the sum formula of \ce{Cu_{1.29(3)}Cr2Te4}. In contrast to the \ce{CuCr2Te4} spinel, we identified an additional Cu position with low occupancy, leading to partial occupation of otherwise empty tetrahedral voids in the spinel structure. Similar conclusions had been drawn from PXRD measurements of copper-rich \ce{Cu_{1+\textit{y}}Cr2Te4} spinels earlier.\cite{Lotgering_1971_excess} The details of our SXRD analysis are given in the SI. EDS measurements on eight crystals of the same batch indicated a sum formula of \ce{Cu_{1.26(9)}Cr_{2.08(8)}Te4}, which is in good agreement with the sum formula based on the SXRD experiment. Using temperature-dependent measurements of the magnetic moment, we identified a ferromagnetic transition at 284\,K, which is slightly lower than the expected $T_{\mathrm{C}}$ of approximately 300\,K determined in previous investigations.\cite{Lotgering_1971_excess} PXRD experiments on crushed crystals confirm the bulk presence of a copper rich spinel. However, it should be noted that \ce{CuCr2Te4} and copper-rich spinels can hardly be differentiated by PXRD as shown by a comparison of their simulated powder patterns which are presented in the Supplementary Information (SI). 

The product of flux-reactions with a lower relative Cu-content (4:1:8) was identified as a mixture of \ce{CuCr2Te4} or \ce{Cu_{1+\textit{y}}Cr2Te4}, CuTe,  as well as a NiAs-type \ce{Cr_{10.7}Te16} phase using PXRD and SXRD experiments. Diffraction and EDS analyses could indicate a doping of \ce{Cr_{10.7}Te8} with Cu, which is further discussed in the SI. %Since the lower Cu-content in the flux led to the formation of a phase that does not contain Cu, further lowering of the Cu-content in the flux would likely only result in a higher percentage of undesired Cr--Te binary phases and was not attempted. The binary Cu--Te phase diagram gives the upper reasonable Cu-content in the flux.\cite{thaddeus1990binary} At a Te content of around 50\,mol\%, the melting temperature of the Cu/Te-flux drastically increases. Based on a centrifugation temperature of 750\,°C, the highest Cu-content, at which the flux is still liquid, is approximately 56\,mol\% Cu. Assuming \ce{CuCr2Te4} is formed by lowering the Cu and Te content, the flux with the highest conceivable Cu content is 7.63:1:8. This ratio is close to the investigated 6:1:8 ratio, when accounting for some cooling until the start of the centrifugation process. Based on these findings, further stoichiometric control of the flux likely would not produce \ce{CuCrTe2}. 
Although metal-flux syntheses have proven to be a useful synthesis strategy for large crystals of the \ce{CuCr2Te4} spinel, our experiments did not result in the formation of \ce{CuCrTe2}.

\textbf{Solvothermal cation exchange reactions:} 
EDS measurements on solid-state synthesized \ce{K_{1--$x$}CrTe2} suggest an understoichiometric K-content with an average composition of \ce{K_{0.74(5)}Cr_{1.037(22)}Te_{2.00(5)}}. It should be noted that the composition might be influenced by the holding times or heating rates, which were not specified in detail by Freitas \textit{et al.}.\cite{Freitas2015}\\
Solvothermal cation exchange reactions between solid-state synthesized \ce{K_{1--$x$}CrTe2} ($x \approx 0.3$) and \ce{CuBr} in acetonitrile at 200\,°C yielded the \ce{CuCr2Te4} spinel. Upon decreasing the synthesis temperature to 90\,°C for one week, careful inspection of the obtained PXRD pattern revealed small but significant differences relative to the simulated pattern of \ce{CuCr2Te4} and all previous synthesis attempts, especially in the region of 45--46° 2$\theta$. SXRD experiments, discussed in detail in the crystal structure section, identified the obtained crystals as \ce{Cu\textsubscript{0.68(3)}CrTe2}. These findings suggest that \ce{Cu\textsubscript{1--$x$}CrTe2} can only be synthesized at low temperatures below 200\,°C. Cation exchange reactions at 90\,°C were also successful with larger flux-grown \ce{K_{1--$x$}CrTe2} crystals, yielding X-ray pure \ce{Cu\textsubscript{1--$x$}CrTe2} crystals with diameters up to 7\,mm. Attempts to perform cation exchange reactions using \ce{LiCrTe2} at 90\,°C yielded unreacted \ce{LiCrTe2} and possibly deintercalated \ce{CrTe2} (see additional information in the SI), demonstrating that the successful synthesis of \ce{Cu\textsubscript{1--$x$}CrTe2} is not only temperature-dependent but also precursor-dependent. The larger interlayer distance in \ce{K_{1--$x$}CrTe2} compared to \ce{LiCrTe2} appears to favor a topochemical reaction over a significant structural rearrangement.

\subsection*{Crystal structure at room-temperature}

\begin{figure*}
 \centering
 \includegraphics[width=0.98\textwidth]{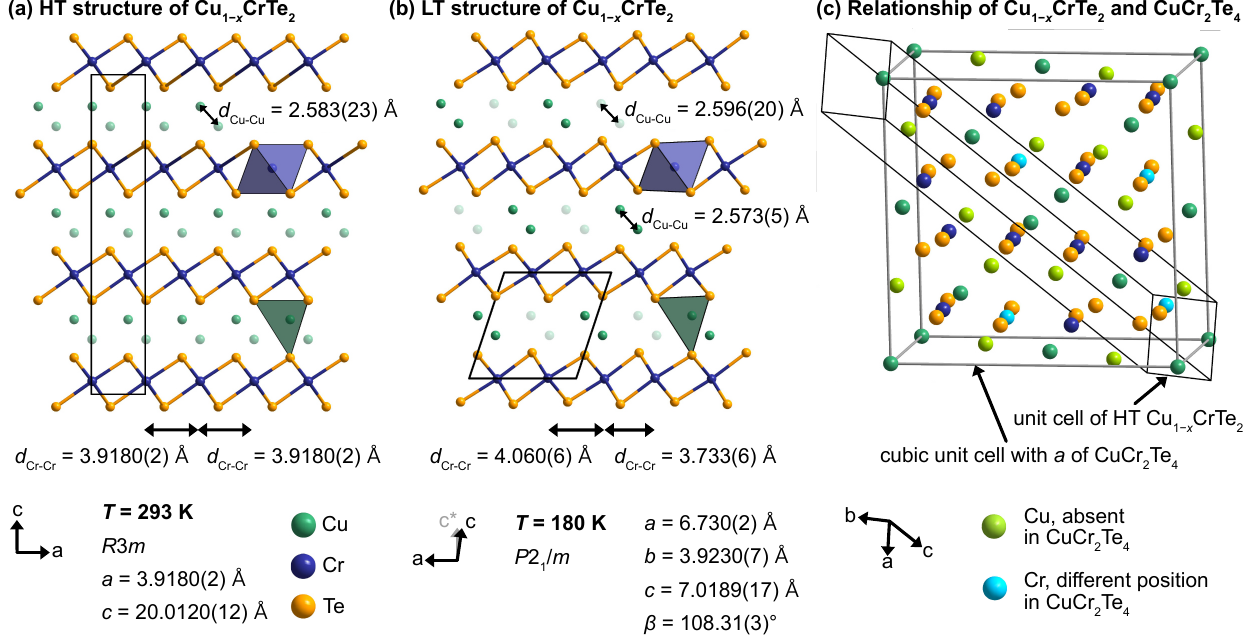}
 \caption{(a) Crystal structure of HT-\ce{Cu_{0.68(3)}CrTe2} at $T$ = 293\,K viewed along the \textbf{b} axis. Cu atoms are colored green, Cr blue, and Te orange. Polyhedra show the octahedral coordination of Cr (blue) and the tetrahedral coordination of Cu (green). (b) Monoclinic distorted crystal structure of LT-\ce{Cu_{0.635(16)}CrTe2} at $T$ = 180\,K viewed along the \textbf{b} axis. Transparency of the Cu atoms is set to $1-2\cdot(\mathrm{S.O.F.})$. (c) Similarities of the crystal structures of HT-\ce{Cu_{0.68(3)}CrTe2} and \ce{CuCr2Te4}. Dark lines show the unit cell of HT-\ce{Cu_{0.68(3)}CrTe2} and grey lines a cubic unit cell with \textit{a}\,=\,11.24(4)\,\AA\,similar to the unit cell of \ce{CuCr2Te4}. Cu atom positions present in HT-\ce{Cu_{0.68(3)}CrTe2} but not \ce{CuCr2Te4} are colored light-green. Cr atom positions that differ between the two crystal structures are colored light-blue. Cu2 atoms in (c) are omitted for clarity.}
 \label{fig:Structure}
\end{figure*}

The crystal structure of \ce{Cu_{1--$x$}CrTe2} was solved from single-crystal X-ray diffraction experiments on \ce{Cu_{1--$x$}CrTe2} obtained from solid-state synthesized \ce{K_{1--$x$}CrTe2} ($x \approx 0.3$). At 293\,K, \ce{Cu_{0.68(3)}CrTe2} crystallizes in the rhombohedral space group $R3m$ with the lattice parameters \textit{a}\,=\,3.9180(2)\,\AA, \textit{c}\,=\,20.0120(12)\,\AA, and \textit{V}\,=\,266.04(3)\,\AA$^{3}$. More detailed information on data collection and refinement is listed in Table \ref{tbl:crydata} and the atomic positions are given in Table \ref{tbl:positions}. As depicted in Figure \ref{fig:Structure}, the crystal structure of \ce{Cu\textsubscript{0.68(3)}CrTe2} is built from \ce{CrTe2} layers, oriented parallel to (001), which are intercalated by Cu atoms. The \ce{CrTe2} layers consist of edge-sharing \ce{CrTe6} octahedra and follow an ABC-stacking motif. Cu atoms in the interlayer space are coordinated in a distorted tetrahedral geometry, with one Te atom being significantly closer than the other three, which belong to the neighboring layer.
The Cu atoms are distributed over partially occupied crystallographic sites with site occupation factors (S.O.F.) of 0.28(2) and 0.40(2), leading to a sum formula of \ce{Cu\textsubscript{0.68(3)}CrTe2}. Apart from the understoichiometry, \ce{Cu\textsubscript{0.68(3)}CrTe2} appears isostructural to other reported \textit{M}Cr\textit{X}\textsubscript{2} (\textit{M} = Cu, Ag, Tl; \textit{X} = S, Se).\cite{Hahn_1957_MCrCh2_Stukturen,Rosenberg_1982_TlCrCh2}
% In the fully stoichiometric structural relatives, the intercalated species show no positional disorder across multiple crystallographic sites, suggesting that the significant Cu understoichiometry in \ce{Cu_{0.68(3)}CrTe2} enables the observed splitting over two sites.

Simulated PXRD patterns based on the structural models of \ce{Cu\textsubscript{0.68(3)}CrTe2} and \ce{CuCr2Te4} (ICSD: 625828) show a degree of similarity (see SI), motivating further investigation of a structural relationship between them.\\
Figure \ref{fig:Structure}(c) depicts a section of the crystal structure of \ce{Cu\textsubscript{0.68(3)}CrTe2}. The unit-cell edges of \ce{Cu\textsubscript{0.68(3)}CrTe2} are colored black. The grey-colored lines represent a cubic unit cell with $a$\,=\,11.24(4)\,\AA, which is just slightly larger than the reported \ce{CuCr2Te4} cell parameter of $a= 11.14$\,\AA. The cubic unit cell has been drawn between the Cu1 positions. 

The basis vectors of the cubic spinel cell ($\mathbf{a}_\mathrm{c}$, $\mathbf{b}_\mathrm{c}$, $\mathbf{c}_\mathrm{c}$) are related to those of the delafossite \ce{Cu_{1--$x$}CrTe2} cell (\textbf{a}, \textbf{b}, \textbf{c}) by:

\begin{equation}
\begin{pmatrix}
\mathbf{a} \\
\mathbf{b} \\
\mathbf{c}
\end{pmatrix}
=
\begin{pmatrix}
0 & -\tfrac{1}{4} & 1 \\
\tfrac{1}{4} & -\tfrac{1}{4} & -1 \\
\tfrac{1}{4} & 0 & 1
\end{pmatrix}
\begin{pmatrix}
\mathbf{a_\mathrm{c}} \\
\mathbf{b_\mathrm{c}} \\
\mathbf{c_\mathrm{c}}
\end{pmatrix}
\end{equation}

Both crystal structures can be derived from a cubic close packing of Te atoms, and accordingly the Te positions in this idealized cubic cell of \ce{Cu_{0.68(3)}CrTe2} closely match the cell of \ce{CuCr2Te4}. The two structures differ, however, in how the octahedral and tetrahedral voids within this packing are occupied. In \ce{CuCr2Te4}, $\frac{1}{2}$ of the octahedral voids are filled with Cr, forming a 3D network through edge-sharing, and $\frac{1}{8}$ of the tetrahedral voids contain Cu. In \ce{Cu_{0.68(3)}CrTe2}, octahedral coordinated Cr form \ce{CrTe2} layers through edge-sharing, whereas Cu occupies tetrahedral voids in the interlayer space. Additional Cu positions (shown in light green in Figure \ref{fig:Structure}(c)) are present in \ce{Cu_{0.68(3)}CrTe2} that have no counterpart in the spinel. Because the Cr and Cu sites cannot be fully mapped between the two structures, no group-subgroup relationship can be established.

\subsection*{Crystal structure at 180 K}

SXRD measurements at $T=$ 180\,K on the same single crystal reveal a monoclinic distortion of the unit cell ($a = 6.730(2)$\,\AA, $b = 3.9230(7)$\,\AA, $c = 7.0189(17)$\,\AA, $\beta$ = 108.31(3)°, $V = 175.93(8)$\,\AA$^3$, space group $P2_{1}/m$ (11)), corresponding to a symmetry reduction from the trigonal $R3m$ high-temperature structure. The $C$-centering expected from a standard trigonal-to-monoclinic descent is absent. This symmetry loss is the result of a disorder-order transition of the Cu atoms in the interlayer space: at $T$ = 180\,K, the Cu1 site refines to a S.O.F. close to 0.5 and was fixed there. A higher occupation factor is unlikely due to close contact between two symmetry-equivalent Cu(I) positions. The Cu2 site has a significantly lower occupancy of 0.135(15). Similar order-disorder transitions have been reported for \ce{CuCrSe2} at $T =$ 365\,K and for \ce{CuCrS2} at $T =$ 688\,K.\cite{Rahman_2025_OrderDisorder,Gagor_2014_CuCrSe2} The transition to the low-temperature modification coincided with a (threefold) twinning by the former threefold rotation axis.

The structural transition also breaks the symmetry of the Cr sublattice. At $T = 293$\,K, the Cr atoms are equidistantly spaced in a triangular lattice ($d$\textsubscript{Cr--Cr} = $3.9180(2)$\,\AA), whereas at $T$ = 180\,K three distinct Cr--Cr distances emerge: $d$\textsubscript{Cr--Cr} = $3.733(6)$\,\AA, $d$\textsubscript{Cr--Cr} = $3.9230(7)$\,\AA\,and $d$\textsubscript{Cr--Cr} = $4.060(6)$\,\AA. As discussed in the magnetic properties section (see below), this distortion is directly linked to the antiferromagnetic spin arrangement. The Cu--Cu distances remain comparable between $293$\,K and $180$\,K, ranging from $2.573(5)$ to $2.596(20)$\,\AA, which is slightly longer than the Cu--Cu distance of $2.304(16)$\,\AA\,reported for disordered \ce{CuCrSe2}.\cite{Gagor_2014_CuCrSe2}

\begin{table}[h]
\small
  \caption{Crystallographic sites in \ce{Cu_{1--$x$}CrTe2} ($x \approx0.3$) including site occupation factors (S.O.F.) and Wyckoff positions at 293\,K in $R3m$ (160) and at 180\,K in ($P2_{1}/m$ (11).The respective structural parameters are listed in Table \ref{tbl:crydata}. Data for $T=100$\,K is listed in th SI.}
  \label{tbl:positions}
  \begin{tabular*}{0.48\textwidth}{@{\extracolsep{\fill}}llllll}
    \hline
    \multicolumn{6}{l}{\textbf{$T$\,=\,293\,K}}\\
    \hline 
    Site  & Wyckoff & S.O.F. & \textit{x} & \textit{y} & \textit{z} \\
    \hline \hline
    Cu1 & 3\textit{a} & 0.28(2) & 2/3 & 1/3 & 0.505(2)\\
    Cu2 & 3\textit{a} & 0.40(2) & 1/3 & 2/3 & 0.5673(12)\\
    Cr1 & 3\textit{a} & 1 & 2/3 & 1/3 & 0.3741(8)\\ 
    Te1 & 3\textit{a} & 1 & 1/3 & 2/3 & 0.44805(7)\\ 
    Te2 & 3\textit{a} & 1 & 2/3 & 1/3 & 0.62763(15)\\ 
    \hline
    \multicolumn{6}{l}{\textbf{$T$\,=\,180\,K}}\\
    \hline 
    Site  & Wyckoff & S.O.F. & \textit{x} & \textit{y} & \textit{z} \\
    \hline \hline
    Cu1 & 2\textit{e} & 0.5 & 0.8870(9) & 1/4 & 0.4068(7)\\
    Cu2 & 2\textit{e} & 0.135(15) & 0.617(4) & 1/4 & 0.593(3)\\
    Cr1 & 2\textit{e} & 1 & 0.7348(6) & 1/4 & 0.9965(5)\\ 
    Te1 & 2\textit{e} & 1 & 1.0075(2) & 1/4 & 0.7790(2)\\ 
    Te2 & 2\textit{e} & 1 & 0.4915(2) & 1/4 & 0.2432(2)\\ 
    \hline
  \end{tabular*}
\end{table}

\begin{table*}
\small
  \caption{Single-crystal X-ray diffraction data for \ce{Cu_{1--$x$}CrTe2} at 293\,K (trigonal high-temperature phase), 180\,K, and 100\,K (monoclinic low-temperature phase). Crystals were obtained by solvothermal cation exchange of solid-state synthesized \ce{K_{1--$x$}CrTe2} ($x \approx 0.3$). Information on the crystallographic sites is listed in Table \ref{tbl:positions}. Data at the three different temperatures was collected on the same single crystal.}
  \label{tbl:crydata}
  \begin{tabular*}{\textwidth}{@{\extracolsep{\fill}}llll}
  %\begin{tabular*}{\textwidth}{@{\extracolsep{\fill}}lll}
    \hline
    \multicolumn{4}{l}{\textbf{Physical, crystallographic, and analytical data}}\\
    \hline
    & \textbf{HT-\ce{Cu\textsubscript{0.68(3)}CrTe2}} & \textbf{LT-\ce{Cu\textsubscript{0.635(16)}CrTe2}} & LT-\textbf{\ce{Cu\textsubscript{0.637(19)}CrTe2}}\\
    Chemical formula &  \ce{Cu\textsubscript{0.68(3)}CrTe2} & \ce{Cu_{0.635(16)}CrTe2} & \ce{Cu_{0.637(19)}CrTe2}\\
    CCDC Deposition code & 2554011 & 2554012 & 2554014\\
    Mol. wt. (g$\cdot$mol$^{-1}$) & 350.38 & 347.55 & 347.55\\ 
    Cryst. syst. & trigonal & monoclinic & monoclinic\\
    Space group & $R3m$ (160) & $P2_{1}/m$ (11) & $P2_{1}/m$ (11) \\
    $a$ (\AA) & 3.9180(2) & 6.730(2) & 6.6599(8)\\ 
    $b$ (\AA) & -- & 3.9230(7) & 3.9186(4)\\
    $c$ (\AA) & 20.0120(12) & 7.0189(17) & 6.9993(8)\\ 
    $\beta$ (°) & -- & 108.31(3) & 108.275(12)\\ 
    $V$ (\AA$^{3}$) & 266.04(3) & 175.93(8) & 173.45(4)\\ 
    $Z$ & 3 & 2 & 2\\
    Calculated density (g$\cdot$cm$^{-3}$) & 6.561 & 6.561 & 6.655\\
    Temperature (K) & 292.99(12) & 180.00(10) & 99.9(2)\\
    Diffractometer & Rigaku XtaLAB Synergy-S & Rigaku Oxford Diffraction SuperNova & Rigaku XtaLAB Synergy-S\\
    Radiation ($\lambda$) & Mo K$\alpha$ (0.71073 \AA) & Mo K$\alpha$ (0.71073 \AA) & Mo K$\alpha$ (0.71073 \AA) \\
    Crystal color & black & black & black\\
    Crystal description & plate & plate & plate\\
    Crystal size (mm$^{3}$) & 0.13 $\times$ 0.11 $\times$ 0.02 & 0.13 $\times$ 0.13 $\times$ 0.03 & 0.13 $\times$ 0.13 $\times$ 0.03\\
    Linear absorption coefficient (mm$^{-1}$) & 23.026 & 22.952 & 23.280\\
    Scan mode & $\omega$ & $\omega$ & $\omega$\\
    Recording range $\theta$ (°) & 3.054 to 30.539 & 3.057 to 30.578 & 2.9610 to 30.7790\\
    $h$ range & --5 $\le h \le$ +5 & --9 $\le h \le$ +9 & --9 $\le h \le$ +9\\
    $k$ range & --5 $\le k \le$ +5 & --5 $\le k \le$ +5 & --5 $\le k \le$ +5\\
    $l$ range & --28 $\le l \le$ +28 & --10 $\le l \le$ +10 & --10 $\le l \le$ +10\\
    Nr. of measured reflections & 2456 & 3140 & 5441\\
    [1ex]\\
    \multicolumn{3}{l}{\textbf{Data reduction}} \\
    Completeness (\%) & 100 & 100 & 100\\
    Nr. of independent reflections & 262 & 612 & 602\\
    $R_{\mathrm{int}}$ (\%) & 3.45 & 5.19 & 5.16\\
    $R_{\mathrm{\sigma}}$ (\%) & 1.08 & 3.26 & 1.85\\
    Absorption correction & numerical (Gaussian grid) & numerical (Gaussian grid) & numerical (Gaussian grid)\\
    Independent reflections \\ with $I$ $\geq$ 2.0$\sigma$ & 262 & 505 & 562\\
    [1ex]\\
    \multicolumn{4}{l}{\textbf{Refinement}}\\
    $R1$ (obs / all) (\%) & 4.45/4.45 & 7.21/8.43  & 8.36/8.61 \\
    $wR2$ (obs / all) (\%) & 11.91/11.91 & 18.23/19.13 & 22.50/22.74\\
    $GOF$ & 1.249 & 1.194 & 1.144\\
    Nr. of refined parameters & 17 & 32 & 32\\
    Nr. of restraints & 1 & 0 & 0\\
    Difference Fourier residues ($e^-$\AA$^{-3}$) & --2.95, 3.99 & --4.84, 5.11 & --6.42, 4.61\\
    \hline
  \end{tabular*}
\end{table*}

\subsection*{Microstructure and elemental composition}

SEM images at various magnifications of \ce{Cu_{1--$x$}CrTe2} ($x \approx 0.3$) synthesized both from flux-grown and solid-state synthesized \ce{K_{1--$x$}CrTe2} ($x \approx 0.3$) (see SI) show plate-like crystals with terraced surface features, consistent with the layered character of the crystal structure. The Br content is negligible, with 0.4(5) mol\% and 0.9(12) mol\% respectively, indicating that the washing procedure effectively removed KBr from the product. The remaining elemental compositions of \ce{Cu_{0.70(3)}K_{0.050(20)}Cr_{0.99(4)}Te_{2.00(3)}} and \ce{Cu_{0.75(3)}K_{0.025(15)}Cr_{1.018(23)}Te_{2.00(3)}}, respectively, are in good agreement with SXRD. Both samples contain trace amounts of K from the precursor, corresponding to an exchange of over 93\% and 96\% of the initial K content. Although the K concentration is close to the EDS detection limit and the K and Te lines overlap spectrally, elemental mapping (see SI) confirms a homogeneous distribution of all elements with no evidence of unreacted \ce{K_{1--\textit{x}}CrTe2}.

\subsection*{Thermal stability}

\begin{figure}[h!]
    \centering
    \includegraphics[width=0.48\textwidth]{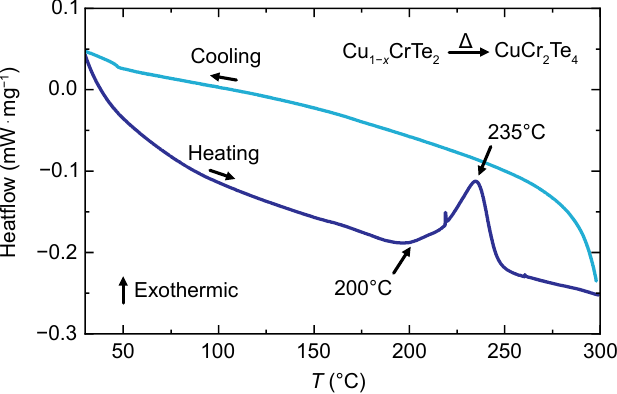}
    \caption{DSC measurement on \ce{Cu_{1--$x$}CrTe2} ($x \approx 0.3$) between RT and 300\,°C. The dark-blue line represents the DSC heat flow signal upon heating and the light-blue line upon cooling. An increasing DSC signal corresponds to an exothermic event, associated with the decomposition of \ce{Cu_{1--$x$}CrTe2}. Arrows point to the onset and maximum of the decomposition signal.}
    \label{fig:DSC_2}
\end{figure}

\noindent To investigate the thermal stability and decomposition of \ce{Cu_{1--$x$}CrTe2} ($x \approx 0.3$) under inert atmosphere, we performed DSC measurements on ground \ce{Cu_{1--$x$}CrTe2} ($x \approx 0.3$) crystals obtained from a solvothermal cation exchange reaction from solid-state synthesized \ce{K_{1--$x$}CrTe2} ($x \approx 0.3$). The temperature program consisted of two cycles: (1) heating and cooling between RT and 150\,°C, and (2) heating and cooling between RT and 300\,°C.\\

During the first heating (and cooling) between RT and 150°C (see SI), no exothermic or endothermic signal was observed, suggesting thermal stability around the employed synthesis temperature of 90\,°C. Upon the second heating, an exothermic event with an onset at $T \approx 200$\,°C and a maximum at $T\approx235$\,°C is observed. Upon cooling to RT no thermal signal was detected, suggesting that the exothermic event at $T \approx 200$\,°C corresponds to an irreversible decomposition. To confirm this, powdered \ce{Cu_{1--$x$}CrTe2} ($x \approx 0.3$) was tempered at 250\,°C for 48\,h in a quartz ampule. PXRD experiments on the tempered powder identified it as \ce{CuCr2Te4} (see SI), confirming the thermal decomposition of \ce{Cu_{1--$x$}CrTe2} ($x \approx 0.3$) to \ce{CuCr2Te4} at approximately 200\,°C. This finding is also in line with our solvothermal cation-exchange reactions at 200\,°C which directly yielded the \ce{CuCr2Te4} spinel. Temperature-dependent measurements of the magnetic moment of \ce{CuCr2Te4} obtained by tempering \ce{Cu_{1--$x$}CrTe2} ($x \approx 0.3$) are depicted in the SI. The sample exhibits a single ferromagnetic transition at $T_{\mathrm{C}} = 327$\,K as previously reported for \ce{CuCr2Te4}.\cite{Suzuyama_2006}

\subsection*{Magnetic properties}

\begin{figure*}[h!]
    \centering
    \includegraphics[width=0.98\textwidth]{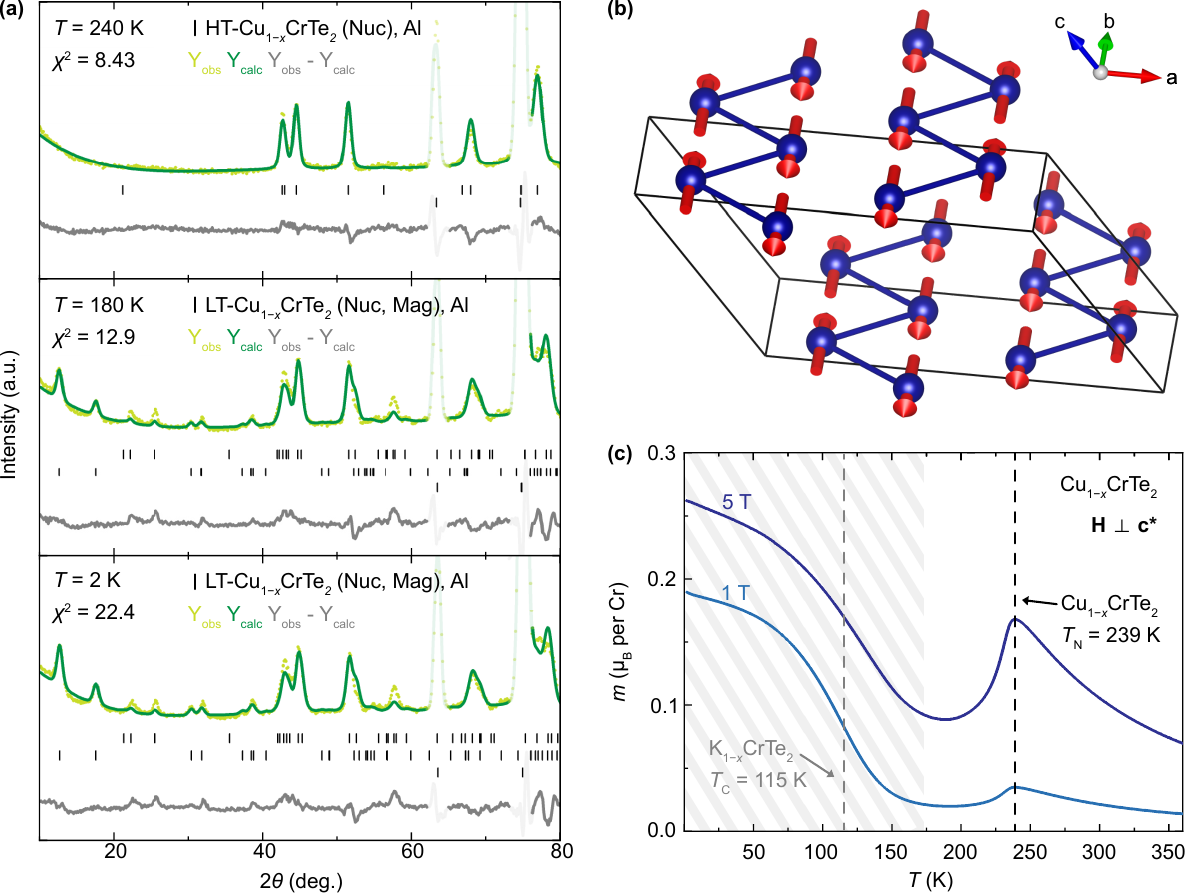}
    \caption{(a) Refined neutron powder diffraction patterns of \ce{Cu_{1--x}CrTe2} ($x \approx 0.3$) collected with a neutron wavelength of 2.449\,\AA\, at $T$ = 240\,K, 180\,K, and 2\,K. Observed intensities are colored light-green, calculated intensities dark-green, differences in grey, and Bragg positions in black. 2$\theta$ intervals of 62--65° and 73--76° are partially transparent for clarity due to strong intensities from the Al container. Full patterns are depicted in the SI. The order of observed phases (left to right) corresponds to the line order of Bragg positions (top to bottom). "Nuc" represents a nuclear only, structural contribution, and "Mag" a purely magnetic contribution. (b) Magnetic structure of \ce{Cu_{1--x}CrTe2} ($x \approx 0.3$). For clarity, only Cr atoms are depicted with red arrows indicating the direction of the magnetic moments. Cr--Cr bonds are depicted to differentiate between closer and more distant Cr atoms. (c) Temperature-dependent measurements of the magnetic moment of \ce{Cu_{1--x}CrTe2} ($x \approx 0.3$) between 1.8\,K and 360\,K for magnetic fields of 1\,T and 5\,T applied perpendicular to \textbf{c*}. }
    \label{fgr:Mag_Neurons}
\end{figure*}

To determine the magnetic structure of \ce{Cu_{1--$x$}CrTe2} ($x \approx 0.3$), neutron powder diffraction (NPD) data were collected, and magnetization measurements were performed. For the NPD, crushed crystals synthesized from flux-grown \ce{K_{1--$x$}CrTe2} ($x \approx 0.3$) were measured at $T =$ 240\,K, 180\,K, and 2\,K. The refined NPD data are shown in \ref{fgr:Mag_Neurons}(a). The 240\,K pattern is well described by the trigonal-rhombohedral high-temperature nuclear structure. At 180\,K, additional reflections appear that cannot be attributed to the monoclinic low-temperature nuclear structure. Using the k-search program in the FullProf suite, these were identified as magnetic reflections described by the propagation vector \textit{k} = (0.5, 0, 0.5). Of the four one-dimensional irreducible representations compatible with this \textit{k}-vector and the $P2_{1}/m$ space group (determined using ISODISTORT; see SI for details), the best fit is obtained with the magnetic space group $P_a2_1/m$ with origin at (0,0,0) (BNS No. 11.55, UNI $P2_1/m.1'_a[P2_1/m]$). From the parent space group $P2_{1}/m$to he MSG $P_a2_1/m$ a \{(2,0,0),(0,-1,0),(-1,0,-1)\} basis transformation was made. The resulting magnetic structure based on the Rietveld refinement at 180\,K is shown in Figure \ref{fgr:Mag_Neurons}(b). At 180\,K, the refined spin arrangement is overall antiferromagnetic. Cr atoms separated by the shorter distance (d\textsubscript{Cr--Cr} = 3.733(6)\,\AA) couple antiferromagnetically, while those at the larger distance (d\textsubscript{Cr--Cr} = 4.060(6)\,\AA) couple ferromagnetically. This is consistent with the Goodenough--Kanamori--Anderson rules:\cite{Anderson1950_GKA,Goodenough1955_GKA,Goodenough1958_GKA,Kanamori1957_GKA_1,Kanamori1957_GKA_2} the shorter Cr--Cr pairs correspond to sharper Cr--Te--Cr angles, favoring antiferromagnetic direct exchange, whereas Cr--Te--Cr angles closer to 90° favor ferromagnetic superexchange. NPD data collected at $T$\,=\,2\,K do not show significant differences compared to those at 180\,K, suggesting no further structural transitions, but resulted in slightly worse refinement parameters. 

Temperature-dependent magnetization measurements (Figure \ref{fgr:Mag_Neurons}(c); full data set in the SI) support the antiferromagnetic ordering. At applied fields of 1\,T and 5\,T, a clear transition is observed at $T$\textsubscript{N} = 239\,K. An additional ferromagnetic-like feature appears near 115--122\,K. This transition closely matches the Curie temperature of the \ce{K_{1--\textit{x}}CrTe2} precursor ($T_{\mathrm{C}} = 116$\,K) and we attribute it to trace amounts of unreacted precursor, whose ferromagnetic signal can dominate even at low concentrations due to its higher net moment, as no change in the long-range magentic order in the NPD data is observed.

\begin{table}[h]
\small
  \caption{Comparison of refined parameters of neutron powder diffraction data on \ce{Cu_{1--$x$}CrTe2} synthesized from flux-grown \ce{K_{1--$x$}CrTe2} ($x \approx 0.3$) collected at $T$ = 240\,K, 180\,K and 2\,K. ccc}
  \label{tbl:neutronsum}
  \begin{tabular*}{0.48\textwidth}{@{\extracolsep{\fill}}llll}
    Temperature (K) & 240 & 180 & 2\\
    \hline
    Space group & $R3m$ & $P_a2_1/m$ & $P_a2_1/m$\\
    \textit{a} (\AA) & 3.9399(4) & 13.4137(23) & 13.3712(26)\\
    \textit{b} (\AA) & -- & 3.9503(10) & 3.9446(11)\\
    \textit{c} (\AA) & 20.030(5) & 8.0436(22) & 8.0291(23)\\
    $\beta$ (°) & -- & 124.201(24) & 124.228(22)\\
    Origin & -- & (0/0/0) & (0/0/0)\\
%    \textit{V} (\AA\textsuperscript{3}) & XXX(X) & XXX(X)& XXX(X)\\
    $m_{b}$ $(\mu_{\rm B})$ & -- & 1.56(6) & 1.89(6) \\
    $R_{p}$ & 6.25 & 6.32 & 6.57\\
    $R_{wp}$ & 8.03 & 8.11 & 8.42\\
    $R_{exp}$ & 2.76 & 2.30 & 1.81\\
    $\chi^{2}$ & 8.43 & 12.5 & 21.5\\
    \hline
  \end{tabular*}
\end{table}

\section*{Conclusions}
We have synthesized \ce{Cu_{1--$x$}CrTe2} ($x \approx 0.3$), the previously missing telluride member of the \ce{CuCr$X$2} series. We find this phase to exist only within a narrow thermodynamic window. The compound is bound by unreacted precursors at low temperatures and decomposes into the structurally related \ce{CuCr2Te4} spinel at temperatures as low as 200\,°C. This narrow stability window renders conventional high-temperature routes ineffective and explains why \ce{CuCrTe2} has remained unreported until now. Nearly phase-pure samples were obtained here by solvothermal cation exchange at 90 °C, preserving the layered topology of the \ce{K_{1--$x$}CrTe2} precursor. 

We found that \ce{Cu_{1--$x$}CrTe2} ($x \approx 0.3$) exhibits robust magnetic order. Neutron powder diffraction reveals an antiferromagnetic ground state below $T_{\mathrm{N}} = 239$\,K, accompanied by a structural transition from a trigonal-rhombohedral to a monoclinic unit cell. The N\'{e}el temperature is notably high within the \ce{$M$CrTe2} family and comparable to the ferromagnetic ordering temperature of fully deintercalated \ce{CrTe2}. The antiferromagnetic spin arrangement obeys the Goodenough--Kanamori--Anderson rules, with the coupling sign governed by the Cr--Te--Cr bond angle. The successful use of solvothermal cation exchange at mild temperatures suggests that similar strategies may provide a route to other missing members of the \ce{$M$Cr$X$2} family and related intercalated van der Waals compounds.

\section*{Author contributions}
FvR designed the experiments. KR and GS synthesized the crystals. KR, GS, FE, VP, and SM conducted the measurements. All authors contributed to the analysis of the data. FvR and KR wrote the manuscript with contributions from all the authors.

\section*{Conflicts of interest}
There are no conflicts to declare.

\section*{Data availability}
Crystallographic data for the LT and HT structures of \ce{Cu_{1--$x$}CrTe2} and \ce{Cu_{1.29}Cr2Te4} have been deposited at the Cambridge Crystallographic Data Centre under deposition numbers CCDC 2554011-2554014. These data can be obtained free of charge from the CCDC via \url{www.ccdc.cam.ac.uk/data_request/cif}. Mcif files of the neutron refinements will be provided in the MAGNDATA database. All other data supporting the findings of this study are available within the article and its Supplementary Information.

\section*{Acknowledgements}
This work was supported by the Swiss National Science Foundation under grants No. PCEFP2\_194183 and No. 200021-204065. Part of this work was performed at the Swiss Spallation Neutron Source (SINQ), Paul Scherrer Institut (PSI), Villigen, Switzerland. The authors would like to thank the UNIGE crystallography service for their help with establishing the structural relationship and Kerry Lee Paglia for conducting the DSC measurement.

\balance

%If notes are included in your references you can change the title from 'References' to 'Notes and references' using the following command:
%\renewcommand\refname{Notes and references}

%%%REFERENCES%%%
\bibliography{rsc} %You need to replace "rsc" on this line with the name of your .bib file
\bibliographystyle{rsc} %the RSC's .bst file

\end{document}